\documentclass[a4paper, oneside, twocolumn, notitlepage, 10pt]{extarticle_ecoc}
\usepackage{ecoc}
\usepackage{amsmath,amsfonts,amssymb}
\usepackage{xcolor}
\definecolor{darkblue}{RGB}{0,0,128}
\definecolor{darkred}{RGB}{128,0,0}
\definecolor{darkgreen}{RGB}{0,128,0}
\definecolor{darkbrown}{RGB}{153,76,0}
\definecolor{boxblue}{RGB}{175,238,238}
\definecolor{dark}{RGB}{175,238,238}
\usepackage{tikz,pgfplots}
\usepackage{pgfplotstable}
\usepackage{flushend}
\newcommand{\eg}{\emph{e.g.,}}

\begin{document}
\selectlanguage{english}    

\title{Few-bit Quantization of Neural Networks for Nonlinearity Mitigation in a Fiber Transmission Experiment}

\author{
    Jamal Darweesh\textsuperscript{(1)}, Nelson Costa\textsuperscript{(2)},
    Antonio Napoli\textsuperscript{(3)}, Bernhard Spinnler\textsuperscript{(3)}, \\ Yves Jaou\"en\textsuperscript{(1)}, Mansoor Yousefi\textsuperscript{(1)}
}

\maketitle

\begin{strip}
 \begin{author_descr}
 
\textsuperscript{(1)} T\'el\'ecom Paris, Palaiseau, France (\textcolor{blue}{\uline{jamal.darweesh@telecom-paris.fr}}) \\ \textsuperscript{(2)} Infinera, Unipessoal Lda, Carnaxide, Portugal ~~ \textsuperscript{(3)} Infinera, Munich, Germany

 \end{author_descr}
\end{strip}

\setstretch{1.1}
\renewcommand\footnotemark{}
\renewcommand\footnoterule{}


\begin{strip}
  \begin{ecoc_abstract}
A neural network  is quantized for the mitigation of nonlinear and components' distortions in a 16-QAM 9x50km dual-polarization fiber transmission experiment. Post-training additive power-of-two quantization at 6 bits incurs a negligible Q-factor penalty. At 5 bits, the model size is reduced by 85\%, with 0.8 dB penalty.

  \end{ecoc_abstract}
\end{strip}

\section{Introduction}

The interaction between the chromatic dispersion (CD), Kerr nonlinearity and amplified spontaneous emission noise 
limits the capacity of optical fiber. Signal processing, such as digital back-propagation (DBP), is  applied at the 
receiver (RX) to mitigate channel impairments.
Neural networks (NNs) have recently been studied for equalization in optical fiber
communication~\cite{zhang2018,butler2021}. 
Compared to model-based equalizers such as  DBP, NNs 
do not require information about the channel, and may offer low-complexity mitigation of impairments \cite{freire2021}.

To implement NNs in electronic receivers, it is necessary to quantize the NN model 
and perform computation in fixed-point arithmetic. In general, weights, biases, activations and 
the data set can be quantized. Quantization and pruning of NNs can reduce the computational complexity 
and memory requirements considerably, while maintaining 
roughly the prediction accuracy.

In this paper, we study several algorithms for the quantization of NNs used for nonlinearity mitigation, in a 16-QAM 34.4 GBaud dual-polarization transmission experiment, over 9 spans of 50km of optical fiber. A low-complexity NN is considered, consisting of two parallel convolutional layers followed by a hidden dense layer, placed after the linear DSP chain at RX. 
We compare post-training, training-aware, additive power-of-two (APoT) \cite{li2019}, uniform, non-uniform, fixed- and mixed-precision quantization, where the convolutional and dense layers are, respectively, quantized at $b_1$ and $b_2\neq b_1$ bits.
Mixed-precision post-training APoT quantization at $b_1=6$ and $b_2=8$ bits is obtained 
with a Q-factor penalty of less than 0.5 dB. The Q-factor begins to drop rapidly below the cut-off values $b_1^c=b_2^c=6$ bits. For instance, at 5 bits, while the model size is reduced by 85\%, the penalty is 0.8 dB.

The Q-factor as a function of the launch power is compared for a number of quantization algorithms and rates. The comparison shows that mixed-precision  training-aware APoT and fixed-precision post-training uniform quantization are, respectively, methods of choice at high and low number of bits (\eg\ $b_{1,2}\leq 8$).

\section{Optical fiber transmission experimental setup}

\begin{figure*}[t]
\centering
\includegraphics[width=0.8\textwidth]{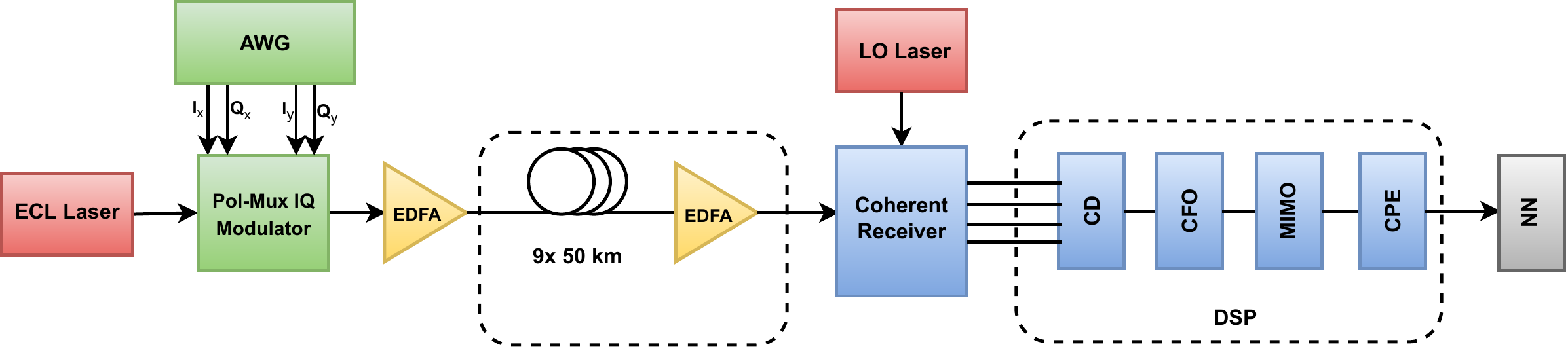}
\caption{Experimental transmission setup.}
\label{fig:sys-model}
\end{figure*}

The fiber-optic transmission experiment setup is shown in Fig.~\ref{fig:sys-model}.
At the transmitter (TX), two sequences of bits  for the $x$ and $y$ polarizations are mapped to two sequences of complex symbols taking values in a 16-QAM constellation, and modulated with root raised cosine (RRC) pulse shape with the roll-off factor 0.1 at $34.4$ GBaud. The two complex-valued digital signals are converted to four continuous electrical waveforms corresponding to the $I$ and $Q$ signals of the $x$ and $y$ polarizations by an arbitrary wave generator (AWG) that includes digital-to-analog converters (DACs) at $88$  Gsamples/s. The electrical signals are converted to optical signals and polarization multiplexed with a Pol-Mux IQ modulator, driven by an external cavity laser (ECL) at wavelength  $1.55~\mu m$ with line width 100 KHz.

The resulting optical signal is sent over a straight-line optical fiber link in a lab, with $9$ spans of Truewave Classic Fiber (TWC) of length $50$ km. 
An Erbium-doped fiber amplifier (EDFA) with 5 dB noise figure is placed at the end of each span to compensate for the fiber loss. The fiber has  $0.23$ dB/km loss, 2.8 ps/(nm-km)
CD, and $2.5$ $({\rm Watt\cdot km)^{-1}}$  nonlinearity parameter. The channel operates in the nonlinear regime at high powers, considering the low dispersion and high fiber nonlinearity parameter; see Fig.~\ref{fig:perf2}.

At the receiver,  the optical signal is polarization demultiplexed and converted to four electrical signals using an integrated coherent receiver, sampled by analog-to-digital converters (ADCs) at  the rate of $50$  Gsamples/s, and equalized  using the conventional linear DSP chain. The linear DSP performs CD compensation,  MIMO equalization (with radius directed equalizer), polarization separation and carrier phase estimation (CPE).  Finally, the resulting signal is passed to a low-complexity NN for the mitigation of nonlinearities and distortions introduced by devices. 

\section{Low-complexity NN for nonlinearity mitigation}

The NN takes four real-valued vectors, corresponding to the real and imaginary parts of the samples of the signals of the $x$ and $y$ 
polarizations, performs nonlinear regression, and outputs two real numbers for each polarization symbol. 

A number of architectures are evaluated. Given the limitations of the practical systems, 
we consider a low-complexity model, with a complex-valued convolutional layer with no activation, processing the signals of the $x$ and $y$ polarizations. 
The complex convolution is implemented using two parallel real-valued filters of length 41.

There are total 82 filter tap weights, far fewer than in generic convolutional layers used in the literature with numerous feature maps. The output of the convolutional layer is  then fed  to a fully-connected layer with $100$ neurons, and tangent hyperbolic (tanh) activation.

Finally, there is an output layer with 2 neurons for each polarization symbol.
Nearest-neighbor symbol detection is applied at the end to detect the symbols of the $x$ and $y$ polarizations.
Note that joint processing of the two polarizations in the dense layer is necessary in order to compensate  nonlinear interactions between the two polarizations during the propagation. 
The NN operates in a sliding-window fashion: as the vector at the input of the 
NN is shifted forward two steps in time, one complex symbol is produced.

\section{Few-bit quantization of the NN equalizer}

The parameters (weights and biases) of the NN, activations and input  data are initially real numbers 
represented in float 32 (FP32), described, \eg\ in the IEEE 754 standards. To implement the NN in hardware 
efficiently, these numbers must be represented by fewer number of bits, \eg\ in INT8 format.
Thus, the real numbers are quantized in a codebook with a finite set of discrete values  
$\mathcal W= \bigl\{0, w^{(1)}, \cdots, w^{(N)}\bigr\}$. The quantization rate of $\mathcal W$ or precision is defined to be $b= \log_2 N $ bits.

\def\sfactor{0.85}

\begin{figure*}[t!]
\begin{center}
\begin{tabular}{c@{~~~~~~~~}c@{~~~~~~~~}c}
\scalebox{\sfactor}{\begin{tikzpicture}

\begin{axis}[scale=0.315,xmin=-4,xmax=4,ymin=3,ymax=9,line width=1, font=\small,
legend entries={\textcolor{black}{NN} , \textcolor{darkred}{Linear DSP}},
legend style={at={(0.65,0.5)}, draw=none, fill=none, legend cell align=left,row sep=2pt},
xlabel={Launch power  [dBm]},
ylabel={Q-factor [dB]},
y label style={at={(0.1,0.5)}},
xtick={-4,-3,-2,-1,0,1,2,3,4,5},
ytick={2,3,4,5,6,7,8,9},
grid,
fill=none,
width=\textwidth
]

\addplot[color=black, smooth, mark=*, mark size=1pt, line width=1.15] file {data/NN.out};
\addplot[color=darkred, dashed, mark=square*, mark size=1pt, line width=1.15] file {data/DSP.out};
\end{axis}
\end{tikzpicture}} &    \scalebox{\sfactor}{ \begin{tikzpicture}

\begin{axis}[scale=0.31,xmin=-4,xmax=4,ymin=4,ymax=9,line width=0.95, font=\small,
legend entries={
\textcolor{black} {Unquantized},
\textcolor{darkblue} {PTQ, $7$ bits} , \textcolor{darkred}{PTQ, $6$ bits}},
legend style={at={(0.63,0.525)}, draw=none, fill=none, legend cell align=left, row sep=2pt},
xlabel={Launch power  [dBm]},
ylabel={Q-factor [dB]},
y label style={at={(0.15,0.5)}},
xtick={-4,-3,-2,-1,0,1,2,3,4,5},
ytick={2,3,4,5,6,7,8,9},
grid,
fill=none,
width=\textwidth
]

\addplot[color=black, smooth, mark=*, mark size=1pt, line width=1.15] file {data/NN.out};

\addplot[color=darkblue, dotted, mark=square*, mark size=1pt, line width=1.1] file {data/NN_post_7.out};

\addplot[color=darkred, dashed, mark=triangle*, mark size=1pt, line width=1.1] file {data/NN_post_6.out};

\end{axis}

\end{tikzpicture}}         &  
\scalebox{\sfactor}{\begin{tikzpicture}

\begin{axis}[scale=0.31,xmin=-4,xmax=4,ymin=4,ymax=9,line width=0.95, font=\small,
legend entries={
\textcolor{black} {Unquantized},
\textcolor{darkblue}{TAQ, $7$ bits},
\textcolor{darkred} {TAQ, $6$ bits}},
legend style={at={(0.64,0.5)}, draw=none, fill=none, legend cell align=left, row sep=2pt},
xlabel={Launch power  [dBm]},
ylabel={Q-factor [dB]},
y label style={at={(0.15,0.5)}},
xtick={-4,-3,-2,-1,0,1,2,3,4,5},
ytick={2,3,4,5,6,7,8,9},
grid,
fill=none,
width=\textwidth
]

\addplot[color=black, smooth, mark=*, mark size=1pt, line width=1.15] file {data/NN.out};

\addplot[color=darkblue, dotted, mark=triangle*, mark size=1pt, line width=1.1] file {data/NN7.out};

\addplot[color=darkred, dashed, mark=square*, mark size=1pt, line width=1.1] file {data/NN6.out};

\end{axis}

\end{tikzpicture}}\\
(a) & (b) & (c) 
\end{tabular}
\end{center}
\caption{a)  Gain of the unquantized NN over linear DSP. Q-factor penalty of b) PT and c) TA quantization with fixed-precision.}
\label{fig:perf2}
\end{figure*}
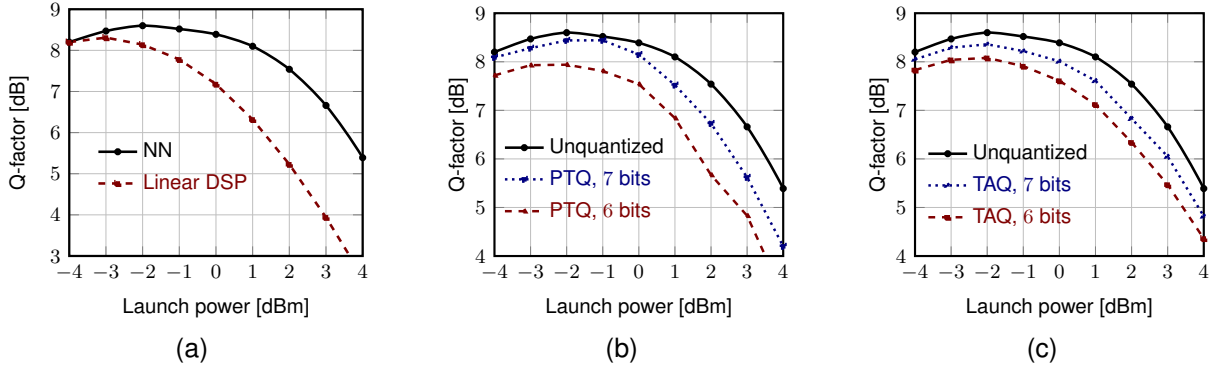

There are two forms of quantization in machine learning. In \emph{post-training quantization (PTQ)}, training is 
performed in full (FP32) or half (FP16) precision. The input tensor, activation outputs, and the resulting weights  are then quantized and used in inference \cite{choukroun}. 
PTQ is fast, but that may come at the expense of accuracy. 

On the other hand, in \emph{training-aware quantization (TAQ)}, quantization  is co-developed with the training algorithm. 
This often results in improved prediction accuracy, because the quantization noise is accounted for \cite{jacob2018quantization}. In this paper, we maximize the Q-factor by searching over a number of TAQ algorithms, notably, the straight-through estimator and several gradient approximation techniques. 
However, TAQ is less suited to real-time processing in high-speed transmission, because it has a higher computational cost than the PTQ (hundred epochs may be required to gain accuracy) and requires hyper-parameter tuning.

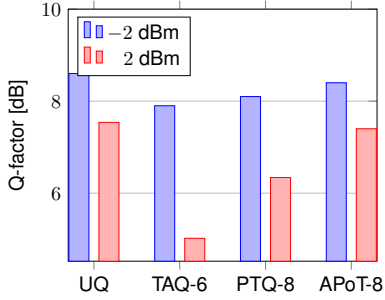
\begin{figure}[h!]
\begin{center}
\scalebox{0.75}{\pgfplotsset{compat=1.15}

\pgfplotstableread[col sep=comma,]{data.csv}\datatable
\begin{tikzpicture}
\begin{axis}[width=7cm,
    ybar=5pt,
    bar width=10pt,
    ylabel={Q-factor [dB]},
    ymax={10},
    xtick=data,
    xticklabels from table={\datatable}{A},
    ymajorgrids,
    legend pos=north west
             ]
    \addplot table [x expr=\coordindex,fill=green, y=B]{\datatable};
    \addplot table [x expr=\coordindex, y=C]{\datatable};
    \legend{$-2$ dBm, $~~~2$ dBm}
\end{axis}
\end{tikzpicture}}
\end{center}
\caption{Comparison of the quantization algorithms, described in the text, at $b_1=6$.}
\label{fig:comp}
\end{figure}

In uniform quantization, the quantization symbols $w^{(i)}$ are placed uniformly between a minimum and maximum weight.

Let $w$ be an unquantized parameter anywhere in the NN,  
$(a,c)$ the smallest interval containing the quantized parameters, $N=|\mathcal W|-1$ and $s(a,c,N){=}(c-a)/(N-1)$.
In uniform quantization, the  quantized weight is $\hat w {=} r s(a,c,N)+a$, where 
$ r {=} \lfloor (c(w,a,c)- a)/s(a,c,N)\rceil$,
$c(w,a,c){=}\min(\max(w,a),c)$ is the clipping function and $\lfloor . \rceil$ is nearest integer.
Quantization is said to be of static range if $a$ and $c$ are known and hard-coded a priori in hardware. 
The same values are used in training and inference, and for all runs. In contrast, in dynamic range quantization, $a$ and $c$ are computed separately for each component of the network.

The distribution of weights of the NN is often Bell shaped. Thus, non-uniform quantization can provide better performance, by assigning more symbols to dense regions. But, non-uniform quantization is not hardware friendly. The power-of-two (PoT) quantization \cite{li2019}
simplifies the implementation by converting multiplications to additions. 
                     
Here, 
\begin{equation*}
\mathcal W (\alpha, b)=  \pm\alpha \Bigl\{0, 2^{0}, 2^{-1}, \cdots, 2^{-(N/2-1)} \Bigr\},
\end{equation*}
where $N=2^b$ and $\alpha$ is stored in FP32, but is applied after the multiply-accumulate operations.
In additive PoT, each quantization symbol is sum of $n$ PoT values, for some $n\in\mathbb N$.
Choose a base number of bits $b_0$ such that $n=b/b_0$ is an integer. Then,
$\mathcal{W'}(\gamma, b) {=}\gamma \sum_{i=0}^{n-1} 2^{-i}\mathcal{W}^n(\alpha, b_0) +\beta $, where $\gamma$ and $\beta$ are scale and shift factors in FP32 that are trainable,
and the set power is per component. It can verified that $|\mathcal W'|=2^b$. The shift parameter $\beta$ allows restricting to unsigned weights.

In mixed-precision quantization \cite{dong2019hawq}, the convolutional layer is quantized at $b_1$ bits and the
dense layers at $b_2\neq b_1$. In fixed-precision, $b_1=b_2$.

\section{Q-factor penalty of quantization algorithms}

Fig.~\ref{fig:perf2} (a) shows the Q-factor gain of the unquantized NN over linear DSP. The gain results in part from the mitigation of dual-pol nonlinearities, and is roughly equal to the DBP gain with large number of spatial steps \cite{freire2021}. The Q-factor penalty of PTQ with fixed precision $b_1=b_2$ is presented in  Fig.~\ref{fig:perf2} (b). PTQ at 6 bits results in a Q-factor drop of $0.7$ dB at $-2$ dbm and $1.9$ dB at $2$ dbm. It can be seen that the quantization penalty increases with the transmission power. TAQ improves the performance, reducing the Q-factor drop to $0.5$ dB at $-2$ and $1.2$ dB at $2$ dBm, as shown in Fig.~\ref{fig:perf2} (c). We compare three quantization algorithms in Fig.~\ref{fig:comp}.
Here, the blue and red bars represent Q-factors at lunch power -2 and 2 dBm, respectively. 
The baseline is the Q-factor of the unquantized (UQ) NN.

In the quantization scheme TAQ-6, uniform fixed-precision TAQ is applied at $b_1=b_2=6$ bits. PTQ-8 corresponds to uniform mixed-precision PTQ, with  $b_1=6$ bits for the convolutional layers and $b_2=8$ bits for the dense. PTQ-8 outperforms TAQ-6, as the Q-factor drop compared to the unquantized NN is reduced to $0.3$ dB at $-2$ dBm and $0.34$ dB at $2$ dBm. Although $b_2=8$ in PTQ-8 compared to 6 in TAQ-6, PTQ-8 is a compelling solution since quantization is done offline after the training. Considering the bell-shaped distribution of the weights of the dense layer, it makes sense to assign more quantization symbols around the mean. 
APoT-8 corresponds to APoT mixed-precision PTQ, with $b_1=6$ bits for the convolutional layers and $b_2=8$ bits for the dense layer.

APoT-8 yields the best performance, with a Q-factor penalty of $< 0.2$ dB at $-2$ and $2$ dBm. Further, APoT-8 has the lowest complexity, since multiplications are implemented by additions in APoT quantization.


\section{Conclusions}

We compared post-training, training-aware, additive power-of-two, uniform, non-uniform, fixed- and 
mixed-precision quantization of the NNs used for nonlinearity mitigation.
A NN is quantized at 6 bits/weight with a Q-factor penalty of $<$0.5 dB, in a dual-pol fiber-optic 
transmission experiment.

\section{Acknowledgements}

This work has received funding from the European Union's Horizon 2020 research and innovation programme under the Marie Sklodowska-Curie grant agreement No. 813144.

\printbibliography

@conference{zhang2018, 
author={Zhang, Shaoliang and Yaman, Fatih and Mateo, Eduardo and Inada, Yoshihisa},

  booktitle={2018 European Conference on Optical Communication (ECOC)}, 

  title={Neuron-Network-Based Nonlinearity Compensation Algorithm}, 

  year={2018},

  volume={},

  number={},

  pages={1-3},
  doi={10.1109/ECOC.2018.8535376}
  }

@article{freire2021, 
author={Freire, Pedro J. and Osadchuk, Yevhenii and Spinnler, Bernhard and Napoli, Antonio and Schairer, Wolfgang and Costa, Nelson and Prilepsky, Jaroslaw E. and Turitsyn, Sergei K.},

  journal={Journal of Lightwave Technology}, 

  title={Performance Versus Complexity Study of Neural Network Equalizers in Coherent Optical Systems}, 

  year={2021},

  volume={39},

  number={19},

  pages={6085-6096},
  
  doi={10.1109/JLT.2021.3096286}
  }

@article{butler2021, 
 author={Butler, Rick M. and Hager, Christian and Pfister, Henry D. and Liga, Gabriele and Alvarado, Alex},

  journal={Journal of Lightwave Technology}, 

  title={Model-Based Machine Learning for Joint Digital Backpropagation and PMD Compensation}, 

  year={2021},

  volume={39},

  number={4},

  pages={949-959},
  doi={10.1109/JLT.2020.3034047}
  }

@article{li2019,
  title={Additive powers-of-two quantization: An efficient non-uniform discretization for neural networks},
  author={Li, Yuhang and Dong, Xin and Wang, Wei},
  journal={arXiv:1909.13144},
  year={2019}
}

@inproceedings{jacob2018quantization,
  title={Quantization and training of neural networks for efficient integer-arithmetic-only inference},
  author={Jacob, Benoit and Kligys, Skirmantas and Chen, Bo and Zhu, Menglong and Tang, Matthew and Howard, Andrew and Adam, Hartwig and Kalenichenko, Dmitry},
  booktitle={Proceedings of the IEEE conference on computer vision and pattern recognition},
  pages={2704--2713},
  year={2018}
}

@INPROCEEDINGS{choukroun,

  author={Choukroun, Yoni and Kravchik, Eli and Yang, Fan and Kisilev, Pavel},

  booktitle={2019 IEEE/CVF International Conference on Computer Vision Workshop (ICCVW)}, 

  title={Low-bit Quantization of Neural Networks for Efficient Inference}, 

  year={2019},

  pages={3009-3018},
   doi={10.1109/ICCVW.2019.00363}
  }

@inproceedings{dong2019hawq,
  title={Hawq: Hessian aware quantization of neural networks with mixed-precision},
  author={Dong, Zhen and Yao, Zhewei and Gholami, Amir and Mahoney, Michael W and Keutzer, Kurt},
  booktitle={Proceedings of the IEEE/CVF International Conference on Computer Vision},
  pages={293--302},
  year={2019}
}

\flushend

\end{document}